\documentstyle[epsfig,12pt]{article}  
\newcommand{\beq}{\begin{equation}}
\newcommand{\eeq}{\end{equation}}
\newcommand{\bea}{\vspace{0.25cm}\begin{eqnarray}}
\newcommand{\eea}{\end{eqnarray}}

\newcommand{\r}{\mbox{{\boldmath
$\rho$}}}

\newcommand{\rb}{\mbox{{\bf
r}}}
\newcommand{\kb}{\mbox{{\bf
k}}}

\newcommand{\pbt}{\mbox{{\bf
p}}_\perp}

\newcommand{\qbt}{\mbox{{\bf
q}}_\perp}

\newcommand{\Abt}{\mbox{{\bf
A}}_\perp}

\newcommand{\Fb}{\mbox{{\bf
F}}}

\newcommand{\fb}{\mbox{{\bf
f}}}

\def\lsim{\mathrel{\rlap{\lower4pt\hbox{\hskip1pt$\sim$}}
    \raise1pt\hbox{$<$}}}         
\def\gsim{\mathrel{\rlap{\lower4pt\hbox{\hskip1pt$\sim$}}
    \raise1pt\hbox{$>$}}}         
\begin{document}
\thispagestyle{empty}
\vspace*{-2cm}
 
\bigskip
 
\begin{center}

  {\large\bf
COLLINEAR PHOTON EMISSION FROM THE QUARK-GLUON PLASMA:
THE LIGHT-CONE PATH INTEGRAL FORMULATION
\\
\vspace{1.5cm}
  }
\medskip
  {\large
  P. Aurenche$^{1}$ and B.G. Zakharov$^{2}$}
  \bigskip

{\it
$^{1}$
LAPTH, Universit\'e de Savoie, CNRS, 9 Chemin de Bellevue, B.P. 110,
F-74941 Annecy-le-Vieux Cedex, France\\
$^{2}$L.D. Landau Institute for Theoretical Physics,
        GSP-1, 117940,\\ Kosygina Str. 2, 117334 Moscow, Russia
\vspace{2.7cm}\\}

  {\bf
  Abstract}
\end{center}
{
\baselineskip=9pt
We give a simple physical derivation of the 
photon emission rate from the weakly coupled quark-gluon
plasma connected with the collinear processes $q\rightarrow \gamma q$
and $q\bar{q}\rightarrow \gamma$. The analysis is based on the
light-cone path integral approach to the induced radiation.
Our results agree with that by Arnold, Moore and Yaffe obtained using
the real-time thermal perturbation theory.
It is demonstrated that the solution of the AMY integral equation
is nothing but the time-integrated Green's function of the  
light-cone path integral approach
written in the momentum representation. 
}

\vspace{1.7cm}

\pagebreak
\newpage

\noindent {\bf 1}.
In recent years much attention has been given to the problem of
the photon radiation from the hot quark-gluon plasma (QGP).
This interest is mostly motivated by the 
possibility of production of the QGP in the heavy-ion collisions 
that are under active investigation at RHIC, and will be studied
soon at LHC. The thermal photon radiation is expected to be a good
probe of formation of the QGP, since the photons can freely escape the
plasma and give information about the plasma temperature, $T$.
The photon emission rate per unit time and volume can be written as
\beq
\frac{dN}{dtdVd\kb}=-\frac{n_{B}(\omega)}{\omega(2\pi)^{3}}
\mbox{Im}\Pi_{\mu}^{\mu}(\omega,\kb)\,,
\label{eq:1}
\eeq
where $\Pi_{\mu}^{\mu}$ is the retarded photon polarization tensor,
$n_{B}(\omega)=(\exp(\omega/T)-1)^{-1}$ is the photon Bose-Einstein factor.
Presently the calculation of the photon polarization
tensor is usually performed within the thermal perturbative Quantum 
Chromodynamics (pQCD) methods. 
The application of the pQCD to description of the QGP includes 
the well-known hard thermal loop (HTL) 
resummation \cite{HTL} which is based on separation of the hard quark and 
gluon modes with momentum $p\sim T$ and soft collective
modes with $p\sim gT$. The HTL approximation leads to the infrared finite
leading order $O(g^{2})$ contribution to the photon radiation rate 
connected with annihilation ($q\bar{q}\rightarrow \gamma g$)  and 
Compton mechanism ($qg\rightarrow \gamma q)$)
\cite{PHOT_BORN}. 

The analysis of the two-loop contribution (connected with 
the bremsstrahlung $q\rightarrow \gamma q$
and the induced annihilation $q\bar{q}\rightarrow \gamma$  
physical processes) to the 
polarization tensor demonstrated \cite{PHOT_2L} that it turns out to be of 
the same order in the coupling constant as the leading order result. 
This fact is a consequence of the collinear nature of the 
bremsstrahlung and induced annihilation, which leads to the divergence of 
their cross sections in the massless limits. In the thermal
bath this divergence is regularized by the thermal quark mass 
$m_{q}^{th} \sim gT$, and as a result these processes give a contribution 
of the same order as the Born approximation.  
It turns out that the higher-order ladder diagrams  
give the contribution of the same order as the two-loop diagrams
\cite{PHOT_LPM}.
The calculation of the ladder diagrams has been performed by 
Arnold, Moore and Yaffe (AMY) \cite{AMY1}. The AMY \cite {AMY1} 
result can be written as 
\bea
\frac{dN}{dtdVd\kb}=\frac{N_{c}\alpha_{em}\sum_{s}q_{s}^{2} }{4\pi^{2}k}
\int\limits_{-\infty}^{\infty}
\frac{dp}{2\pi} n_{F}(p+k)[1-n_{F}(p)]
\frac{(p+k)^{2}+p^{2}}{(p+k)^{2}p^{2}}\nonumber\\
\times
\mbox{Re}
\int \frac{d\pbt}{(2\pi)^{2}}  
\pbt\cdot
\fb(\pbt,p,\kb)\,,
\label{eq:10}
\eea
where
$q_{s}$ is the charge of quark species type $s$,
$n_{F}(p)=1/(\exp(p/T)+1)$ is the thermal Fermi distribution, 
and $\fb$
satisfies the integral equation
\beq
i\delta E\fb(\pbt,p,\kb)=2\pbt-g^{2}C_{F}T
\int\frac{d\qbt}{(2\pi)^{2}} C(\qbt)\,
[\fb(\pbt,p,\kb)-\fb(\pbt-\qbt,p,\kb)]\,.
\label{eq:20}
\eeq
Here, $C_{F}=4/3$ is the quark Casimir, 
$\delta E=|k(\pbt^{2}+m_{q}^{2})/2(p+k)p|$,
$m_{q}$ is the asymptotic thermal quark 
quasiparticle mass, the kernel of the integral equation 
can be expressed via the HTL resummed gluon propagator.
It can be written in a simple form 
as $C(\qbt)=m_{D}^{2}/\qbt^{2}(\qbt^{2}+m_{D}^{2})$ \cite{PA_C}, where
$m_{D}=gT[(N_{c}+N_{F}/2)/3]^{1/2}$ is the Debye mass.
The $C(\qbt)$ includes the contribution of both the screened longitudinal
gluon as well as the unscreened transverse gluon.

Physically the ladder diagrams describe the 
Landau-Pomeranchuk-Migdal (LPM) 
suppression \cite{LP, Migdal} of the collinear bremsstrahlung/annihilation.
In the last decade considerable progress has been made 
in understanding the LPM effect in QCD for fast partons with 
energy $E\gg T$ propagating through the QGP. 
Two different approaches to the LPM effect in this regime have been 
developed: the BDMPS 
formalism \cite{BDMPS,BDMS}
and the light-cone path integral (LCPI) formalism \cite{Z1,Z2,Z_PAN98,Z3}
(for a review, see \cite{BSZ,Z_NP2005}). 
In the LCPI approach the radiation rate is expressed via 
the Green's function of a two-dimensional Schr\"odinger equation 
with an imaginary potential.

In the studies
\cite{BDMPS,BDMS,Z1,Z2} the QGP has been modeled by
a static system of the Debye screened color centers.
In the case of the photon radiation from the thermal partons 
one should treat on the same footing the partons from which the photon
is radiated and the target partons. In the AMY analysis \cite{AMY1}
it was done using the real time thermal perturbation theory.
In the present paper we give an alternative simple physical derivation
of the AMY result within the LCPI approach.
We demonstrate that in the case of the weakly coupled QGP
when one can separate the hard and soft modes the treatment
of the LPM suppression for the photon emission from the 
thermal partons can be made similarly to the case of fast partons.
We show that the solution of the AMY integral equation (\ref{eq:20})
is nothing but the time-integrated Green's function of the LCPI approach
written in the momentum representation. 

Whether the model of the weakly coupled QGP, used in the present and
AMY \cite{AMY1} analyses, is a reasonable 
approximation to the QGP which is produced in $AA$-collisions 
is still an open question.
However, in principle, our derivation may have a broader range
of applicability then the pQCD description of the QGP, if one
calculates somehow the function $C(\qbt)$ and the quark quasiparticle mass 
using the non-perturbative methods, or, say, extracts them fitting 
the experimental data on the jet quenching.

\vspace{.1cm}
\noindent {\bf 2}. 
We consider the radiation of the photons with $\omega \gsim T$ which
is controlled by the dynamics of the hard quark modes in the QGP. 
The average energy for hard partons $\sim 3T$. In the weakly coupled 
QGP these hard partons undergo typically only small angle multiple 
scattering due to interaction with 
the random  gluon fields at the momentum scale $\sim gT \ll T$. 
The large angle
scattering is a very rare process. The typical separation between 
the large angle scatterings is $L_{l.a.}\sim 1/g^{4}T$. Using this scale, 
from the uncertainty relation $\Delta E \Delta t\sim 1$, one can
obtain for the typical off-shellness of the hard modes $\delta m \sim
(T/L_{l.a.})^{1/2}\sim g^{2}T$, which turns out to be much smaller than
the thermal quasiparticle masses for quarks and gluons 
$m_{q,g}^{th}\sim gT$.
Note that the $L_{l.a.}$ at the same time gives the scale at which 
the soft rescatterings generate the scattering angle about unity.
It follows from the $L$-dependence of the mean squared momentum 
transfer $q^{2}\sim g^{4}T^{3}L$ connected with the random walk of
hard partons due to small angle multiple scattering.

The above estimates show that at the space scale considerably smaller
than $L_{l.a}$ the hard parton trajectories are almost straight lines
slightly distorted by soft rescatterings. 
Thus, at $L\ll L_{l.a.}$ the hard modes can be 
described as the quasiclassical plane waves propagating in the 
soft random gluon field.
It is important that in first approximation one can neglect
the statistics effects in treating the small angle scattering.
Indeed, the typical space scale for the soft gluon modes $1/gT$
is much larger than the typical separation between the hard partons
$\sim 1/T$. For this reason from the point of view of the hard parton
scattering the soft gluon field can viewed as a uniform field
at the scale $\sim 1/T$. In the uniform field all hard partons 
of the same type scatter in the same way, and the small angle scattering
leads simply to some shift of the distribution function in the 
momentum space. A well-known example of this phenomenon is shift
of the Fermi-sphere for the strongly degenerated Fermi gas in a uniform
electric field. Any statistics effects will be suppressed by some power of $g$.

The above facts on the hard parton trajectories 
are important 
from the point of view of the treatment of the $q\rightarrow \gamma q$ and 
$q\bar{q}\rightarrow \gamma$ processes in the pQGP. 
Let us
first consider the photon bremsstrahlung. From the uncertainty relation
one gets for the photon formation length 
$L_{f}\sim 2(1-x)E/xm_{q}^{2}$
\footnote{This estimate does not account for the LPM
suppression. If the radiation rate is suppressed by a factor 
$S_{LPM}$, the real in-medium formation length 
is $L_{f}^{LPM}\sim S_{LPM}L_{f}$.},
where $E$ is the 
initial quark energy, and $x=\omega/E$ is the photon fractional 
longitudinal momentum (we choose the $z$ axis along the initial quark
momentum). For $\omega\sim E\sim T$ one obtains $L_{f}\sim 1/g^{2}T$.
Thus, one sees that $L_{f}\ll L_{l.a}$. 
It means that at leading order in
$g$  the photon bremsstrahlung is not affected
by the hard scattering. Since the statistics 
effects can be neglected for small angle scattering as well, 
the bremsstrahlung 
occurs in the same way as for hard partons in a soft random gluon field.
The only effect of the thermal bath of hard partons is the
trivial Pauli-blocking for the final quark, which, of course, 
must be taken into account. 
Due to the collinear character of the induced 
photon emission, in accounting for 
the Pauli-blocking one can take the final
quark momentum collinear to that for the initial quark.
In this case the contribution
of the bremsstrahlung into the photon emission rate 
can be written as
\beq
\frac{dN_{br}}{dtdVd\kb}=\frac{d_{br}}{k^{2}(2\pi)^{3}}
\sum_{s}
\int_{0}^{\infty} dp p^{2}n_{F}(p)[1-n_{F}(p-k)]\theta(p-k)
\frac{dP^{s}_{q\rightarrow \gamma q}(p,k)}{dk dL}\,,
\label{eq:30}
\eeq
where 
$d_{br}=4N_{c}$ is the number of the quark and antiquark states, 
$
{dP^{s}_{q\rightarrow \gamma q}}(p,k)/{dk dL}$ is the probability  
of the photon emission 
per unit length from a fast quark of type $s$ in the random soft gluon field
without the thermal bath.

The formation length for 
the $q\bar{q}\rightarrow \gamma$ mechanism is also much smaller 
than $L_{l.a.}$. It means that the photon is produced 
in annihilation of almost collinear quark-antiquark pair induced
by the small angle rescatterings in the soft random gluon field.
The contribution to the radiation rate can be easily 
obtained using the detailed balance principle. 
For the plasma size much smaller than the photon
absorption length it gives
\beq
\frac{dN_{an}}{dtdVd\kb}=
[1+n_{B}(k)]^{-1}
\frac{dN_{abs}}{dtdVd\kb}\,.
\label{eq:40}
\eeq
The photon absorption rate on the right-hand side of Eq. (\ref{eq:40})
should be evaluated for the thermolized photons. 
It can be written via the 
probability distribution per unit length
for the $\gamma \rightarrow q\bar{q}$ transition
$
{dP^{s}_{\gamma\rightarrow q\bar{q}}(k,p)}/{dp dL}
$, where $p$ is the
final quark momentum, similarly to the formula (\ref{eq:30}). 
Then, using (\ref{eq:40}) one
obtains
\beq
\frac{dN_{an}}{dtdVd\kb}=\frac{d_{an} }{(2\pi)^{3}}
\sum_{s}
\int_{0}^{\infty} dp n_{F}(p)n_{F}(k-p)\theta(k-p)
\frac{dP^{s}_{\gamma\rightarrow q\bar{q}}(k,p)}{dp dL}\,,
\label{eq:50}
\eeq
where $k-p$ is the antiquark momentum, $d_{an}=2$ is the number
of the photon helicities.
 
\vspace{.1cm}
\noindent {\bf 3}. 
The probability distributions entering (\ref{eq:30}) and 
(\ref{eq:50}) for the case of 
the random soft gluon field can be evaluated in the LCPI 
approach \cite{Z1} similarly to the case of the static model of 
the QGP which has been used in \cite{Z1} only for simplicity.
For this reason we limit ourselves to highlighting only the 
aspects connected with the generalization 
to the arbitrary random external gluon field.
The starting point of the LCPI approach is the representation
of the parton wave function in the form 
\beq
\psi(t,\rb)=\exp(-iE\xi)\phi(z,\xi,\r)\,,
\label{eq:60}
\eeq
where $\xi=t-z$, $\rb=(z,\r)$, and it is assumed that $E>>m_{q}$.  
To leading order in $m_{q}/E$ the z-dependence of
the transverse wave function $\phi(z,\xi,\r)$ for fixed $\xi$ is governed
by the two-dimensional Schr\"odinger equation
\beq
i\frac{\partial\phi(z,\r)}{\partial
z}=
\hat{H}\phi(z,\r)\,,
\label{eq:70}
\eeq
with the
Hamiltonian
\beq
\hat{H}=\frac{
[(\pbt-g\Abt)^{2}
+m^{2}_{q}]} {2\mu_{q}}
+g(A_{0}-A_{z})\,.
\label{eq:80}
\eeq
Here, the Schr\"odinger
mass is
$\mu_{q}=E$, $\pbt$ is the operator of transverse
momentum, $A_{\mu}=(A_{0},-\Abt,-A_{z})$ is the external 
gluon vector potential ($A_{\mu}$ should be understood as 
$\hat{\tau}^{a}_{q}A^{a}_{\mu}$, for clarity we omit the color generators).
The argument $\xi$ of $\phi$ is omitted in (\ref{eq:70}), since  the $\xi$ 
dependence emerges only via the boundary condition for $\phi$. 
For a random external potential 
without loss of generality we can set
$\xi=0$. For this choice
the vector potential in the Hamiltonian (\ref{eq:80}) must be 
taken at $x^{\mu}=(t,\r,z)$.
Thus the Hamiltonian (\ref{eq:80}) describes the evolution
of the partonic wave function along the light-cone.
Eqs. (\ref{eq:60}), (\ref{eq:70}) hold for
each helicity state. 
The photon wave function can be written in a similar way (of course,
in this case the interaction and mass terms do not appear).  

The matrix elements for the $q\rightarrow \gamma  q$ 
and $\gamma\rightarrow q\bar{q}$ transitions can be written
in terms of the Green's function for the Hamiltonian (\ref{eq:80})
and its photon analogue.
In the LCPI approach they are written in the Feynman path integral
form. The quark Green's function reads
\beq
K_{q}(\r_{2},z_{2}|\r_{1},z_{1})=
\int {\cal
D}\r
\exp
\left\{
i\int
dz
\left[
\frac{\mu_{q} \dot{\r}^{2}}{2}-g\,
\dot{x}^{\mu}A_{\mu}(\r,z)
-\frac{m_{q}^{2}}{2\mu_{q}}
\right]
\right\}\,\,,
\label{eq:90}
\eeq
where $\dot{\r}=d\r/dz$, $\dot{x}^{\mu}=(1,\dot{\r},1)$.
Since $|\dot{\r}|\ll 1$ one can replace 
$\dot{x}^{\mu}$ by $u^{\mu}=(1,0_{\perp},1)$. 
As compared with Ref. \cite{Z1} 
where the static potential with only $A_{0}(\rb)$ nonzero component
has been used, now it is replaced by 
$A^{-}=u^{\mu} A_{\mu}(x)$ with $x^{\mu}=(z,\r,z)$. 

Let us consider the $q\rightarrow \gamma q$ transition.
The probability of the photon emission per unit length
diagrammatically is given by the graph a of Fig.~1, where the longitudinal
coordinate of one of the vertex is fixed. 
Note that due to the relation 
$-\hat{\tau}_{q}^{*}=\hat{\tau}_{\bar{q}}$ 
the quark Wilson lines in the path integral for quark propagators
turns out to be replaced by the antiquark Wilson lines in 
the complex conjugate quark propagators.

Similarly to the static potential one can 
perform in the diagram 1a averaging over the external 
potential at the level of integrands. Then the 
interaction with the external 
potential turns out to be transformed into the interaction 
between trajectories which depends only on the relative 
transverse distances between the trajectories.
This fact allows one to perform some path integrations
analytically, and transform the graph 1a to 1b. 
(the details can be found in \cite{Z3,BSZ,Z_NP2005}),
in which the only dynamical variable
is the transverse separation between quark and photon 
$\r=\r_{q}-\r_{\gamma}$, and the antiquark is located at the center 
of mass of the $q\gamma$-system.
The propagator describing the evolution of a spurious  
$q\bar{q}\gamma$-system in graph b of Fig.~1
in the path integral form reads
\beq
{\cal K}(\r_{2},z_{2}|\r_{1},z_{1})=
\int {\cal
D}\r
\exp
\left\{
i\int
dz
\left[
\frac{M(x) \dot{\r}^{2}}{2}- \frac{1}{L_{f}}
\right]
\right\}
\Phi(\{\r\},z_{1},z_{2})
\,\,,
\label{eq:100}
\eeq
where $x=\omega/E$ is the photon fractional momentum, 
$M(x)=Ex(1-x)$ is the reduced Schr\"odinger mass,
$L_{f}=2E(1-x)/x m_{q}^{2}$ is the photon formation length
which comes from the mass terms in the quark propagators 
(\ref{eq:90}), and
\beq
\Phi(\{\r\},z_{1},z_{2})= 
{\Large\langle\Large\langle}
\exp\left\{-ig\int dz [A^{-}_{q}(\r_{q},z)
+A^{-}_{\bar{q}}(\r_{\bar{q}},z)]\right\}
{\Large\rangle\Large\rangle}
\label{eq:110}
\eeq
is the averaged Wilson line factor. The quark and antiquark transverse
coordinate in (\ref{eq:110}) in therms of $\r$ are given by $\r_{q}=x\r$,
$\r_{\bar{q}}=0$.
For nonzero $\r_{1,2}$ the propagator (\ref{eq:100}) is not gauge 
invariant without the  
Wilson lines connecting
the quark trajectories at $z_{1,2}$. This fact, however, is not important
since in the final formula for the radiation rate this propagator only appears 
for the point-like initial and final states, i.e., for the configurations
shown in Fig.~1b. The formula for the probability of the photon
emission per unit length is similar to that for the static potential
(we use here the fractional photon momentum $x$ instead of 
$k$ in (\ref{eq:30}))
\beq
\frac{d P_{q\rightarrow \gamma q}^{s}}{d
x dL}=2\mbox{Re}
\left.\int\limits_{-\infty}^{0} d
z
\hat{g}(x)\left[
{\cal K}(\r_{1},0|\r_{2},z)
-{\cal K}_{vac}(\r_{1},0|\r_{2},z)
\right]\right|_{\r_{1,2}=0}\,.
\label{eq:120}
\eeq
where $K_{vac}$ is the free propagator for $\Phi=1$
\footnote{In the case of an infinite plasma the vacuum term in 
(\ref{eq:120}) gives zero contribution. However, it is necessary for
selection of the induced radiation for a hard parton produced in the medium
}.
The factor $\hat{g}$ in (\ref{eq:120}) is the vertex operator 
which accumulates the
summation over the particle helicities. It is given by
\beq
\hat{g}(x)=\frac{1}{2}
\sum_{\{\lambda\}} 
V^{*}(x,\{\lambda\})
V(x,\{\lambda\})
\left(\frac{\partial}{\partial
\rho_{x}}-
i\lambda_{\gamma}\frac{\partial}{\partial
\rho_{y}}\right)^{*}_{2}\cdot
\left(\frac{\partial}{\partial
\rho_{x}}-
i\lambda_{\gamma}\frac{\partial}{\partial
\rho_{y}}\right)_{1}\,,
\label{eq:130}
\eeq
\beq
V(x,\{\lambda\})=
\frac{-iq_{s}}{2 M(x)}\sqrt{\frac{\alpha}{2x}}
\left[\lambda_{\gamma}(2-x)+2\lambda_{q}x\right]\,,
\label{eq:140}
\eeq
where $\{\lambda\}$ is the set of helicities for 
the $q\rightarrow \gamma q$
transition, as in (\ref{eq:10}) $q_{s}$ is the quark charge.
The vertex factor (\ref{eq:130}) corresponds to the averaged 
over the quark color states probability, the corresponding normalization factor
$1/N_{c}$ is canceled by the $N_c$ which comes from the summing
over the colors in the quark loop in the diagram of Fig.~1b. 

Taking the advantage of the fact that the correlation radius for 
soft modes is much 
smaller than the photon formation length, 
the Wilson line factor $\Phi$ can be written in the 
exponential form
\beq
\Phi(\{\r\},z_{1},z_{2})= 
\exp\left[-\int_{z_{1}}^{z_{2}} dz P(x\r(z))\right]\,,
\label{eq:150}
\eeq
where, to leading order in the coupling constant, $P(\r)$ 
diagrammatically is 
given by the graphs a,~b,~c of Fig.~2. The wavy lines in Fig.~2 correspond
to the gluon correlator (the color indexes are omitted)
\beq
G(x-y)= 
u_{\mu}u_{\nu}
{\Large\langle\Large\langle}
A^{\mu}(x)A^{\nu}(y)
{\Large\rangle\Large\rangle}\,.
\label{eq:160}
\eeq
The parallel straight lines in Fig.~2 show the quark trajectories 
along the light-cone
$t=z$, and the $z$-integrations 
along trajectories are performed with a fixed position of one
of the vertex. We neglect  the contribution to the function
$P(\r)$ from the higher order diagrams, some of the diagrams of the order 
$g^{4}$ are shown in Figs.~2d,~e,~f. 
This approximation 
assumes that the interaction of hard partons with soft modes should be 
sufficiently small at the longitudinal scale about the Debye 
screening radius. One can say, that neglecting the higher order 
contribution to $P(\r)$
corresponds to the dilute gas approximation for the gluon correlators.
In this approximation one gets
\beq
P(\r)=g^{2}C_{F}\int\limits_{-\infty}^{\infty} dz 
[G(z,0_{\perp}z)-G(z,\r,z)]\,.
\label{eq:170}
\eeq
Here we have taken into account the effect of the quark/antiquark 
color generators, which leads to appearing of the quark Casimir 
factor for each gluon correlator.
Note that the QED analogue of (\ref{eq:170}) has been introduced in 
\cite{Z_A2e} in the analysis of propagation of the relativistic 
positronium atoms in amorphous media. One can write $P(\r)$
as the momentum integral
\beq
P(\r)=\int \frac{d\qbt}{(2\pi)^{2}} [1-\exp(i\r \qbt)]P(\qbt)\,.
\label{eq:180}
\eeq
One can easily show that if one uses for the gluon 
correlators their HTL expressions
the $P(\qbt)$ can be expressed via the function $C(\qbt)$
which enters (\ref{eq:20}) as
\beq
P(\qbt)=g^{2}C_{F}TC(\qbt)\,.
\label{eq:190}
\eeq
Note that for the static model of the QGP the function $P(\r)$ reads
\beq
P(\r)= 
\frac{n{\sigma}_{q\bar{q}}(\rho )}{2}\,,
\label{eq:200}
\eeq
where $n$ is the number density
of the color centers, and $\sigma_{q\bar{q}}(\rho)$ is 
the dipole cross section
\beq
\sigma_{q\bar{q}}(\rho)={C_{T}C_{F}\alpha_{s}^{2}}\int d\qbt
\frac{[1-\exp(i\qbt\r)]}{(\qbt^{2}+m_{D}^{2})^{2}}\,\,.
\label{eq:210}
\eeq
Here $C_{T}$ is the color Casimir of the scattering center.
It is interesting that numerically the $P(\r)$ 
evaluated in the static model
turns out to be 
close to that calculated with the HTL formula 
for $C(\qbt)$ in the region $\rho \lsim 1/m_{D}$ which is important
for the radiation processes in the QGP.

For the interaction factor written in the form (\ref{eq:150}) the propagator
(\ref{eq:100}) is, evidently, the Green's function for the Hamiltonian
\beq
\hat{\cal{H}}=-\frac{1}{2M(x)}
\left(\frac{\partial}{\partial \r}\right)^{2}
          -iP(\r x) +\frac{1}{L_{f}}\,.
\label{eq:220}
\eeq
The formulas for the radiation rate in terms of the solution of 
the Schr\"odinger equation with the Hamiltonian (\ref{eq:220})
with smooth boundary conditions which are convenient
for numerical calculations can be found in \cite{Z_PAN98,Z_RAA}.

To establish the connection between the above formulas of the 
LCPI approach and those of AMY \cite{AMY1}
we rewrite formula (\ref{eq:120}) in terms of the in-medium 
light-cone wave function 
of the spurious $q\bar{q}\gamma$-system 
$
\Psi_{m}(x,\r,\{\lambda\})
$. 
For $\lambda_{q'}=\lambda_{q}$ it reads
\bea
\Psi_{m}(x,\r,\{\lambda\})=
V(x,\{\lambda\})
\left.\left(\frac{\partial}{\partial
\rho_{x}^{'}}-
i\lambda_{\gamma}\frac{\partial}{\partial
\rho_{y}^{'}}\right)
\int\limits_{-\infty}^{0}dz {\cal
K}(\r,0|\r',z)
\right|_{\r^{'}=0}\,\,.
\label{eq:230}
\eea
The analogue of (\ref{eq:230}) in the vacuum case
after $z$-integration 
gives the well-known
light-cone wave function of the $q\gamma$ Fock component in the quark
\bea
\Psi(x,\r,\{\lambda\})=
V(x,\{\lambda\})
\exp(-i\lambda_{\gamma}\varphi)m_{q}K_{1}(\rho
m_{q}x)\,,
\label{eq:240}
\eea
here $K_{1}$ is the Bessel function.
In terms of $\Psi_{m}$
formula (\ref{eq:120}) reads
\beq
\frac{d P_{q\rightarrow \gamma q}^{s}}{d
x dL}=\mbox{Re}
\sum_{\{\lambda\}}
V^{*}(x,\{\lambda\})
\left.
\left(\frac{\partial}{\partial
\rho_{x}}-
i\lambda_{\gamma}\frac{\partial}{\partial
\rho_{y}}\right)^{*}
\Psi_{m}(x,\r,\{\lambda\})
\right|_{\r=0}\,.
\label{eq:250}
\eeq
The function $\Psi_{m}$ satisfies the integral equation
\beq
\Psi_{m}(x,\r,\{\lambda\})=\Psi(x,\r,\{\lambda\})-
\int d\r' W(\r,\r')P(x\r')\Psi_{m}(x,\r',\{\lambda\})
\label{eq:260}
\eeq
with
\beq
W(\r,\r')=
\int\limits_{-\infty}^{0}dz {\cal
K}_{vac}(\r,0|\r',z)=
\int\limits_{-\infty}^{0}dz 
\frac{M(x)}{2\pi i|z|}
\exp
\left\{i
\left[
\frac{M(x)(\r-\r')^{2}}{2|z|}- \frac{|z|}{L_{f}}
\right]
\right\}
\,.
\label{eq:270}
\eeq
In momentum representation it takes the form
\bea
\Psi_{m}(x,\pbt,\{\lambda\})
=
V(x,\{\lambda\})
\frac{p_{x}-i\lambda_{\gamma}p_{y}}{\pbt^{2}+\epsilon^{2}}
\nonumber\\
+\frac{2i M(x)}{\pbt^{2}+\epsilon^{2}}
\int \frac{d\qbt}{(2\pi)^{2}}P(\qbt)
[\Psi_{m}(x,\pbt,\{\lambda\})
-\Psi_{m}(x,\pbt-x\qbt,\{\lambda\})]
\label{eq:280}
\eea
with $\epsilon=x m_{q}$.
Evidently the $\Psi_{m}$ can be written in the form
\beq 
\Psi_{m}(x,\pbt,\{\lambda\})=V(x,\{\lambda\})
[F_{x}(\pbt)-i\lambda_{\gamma}F_{y}(\pbt)]\,,
\label{eq:290}
\eeq
where now $\Fb$ is the solution to the integral equation
\beq
\Fb(\pbt)
=
\frac{\pbt}{\pbt^{2}+\epsilon^{2}}
+\frac{2i M(x)}{\pbt^{2}+\epsilon^{2}}
\int \frac{d\qbt}{(2\pi)^{2}}P(\qbt)
[\Fb(\pbt)
-\Fb(\pbt-x\qbt)]\,.
\label{eq:300}
\eeq
Making use of (\ref{eq:140}) and (\ref{eq:290}) 
after summing over the helicities 
one can rewrite (\ref{eq:250}) in momentum representation as
\beq
\frac{d P^{s}_{q\rightarrow \gamma q}}{d
x dL}=\frac{2q_{s}^{2}\alpha_{em}[1+(1-x)^{2}]}{xM(x)}\mbox{Im}
\int \frac{d\pbt}{(2\pi)^{2}} \,\pbt\cdot \Fb\,.
\label{eq:310}
\eeq

Similar calculations for $\gamma \rightarrow q\bar{q}$ transition
give
\beq
\frac{d P^{s}_{\gamma\rightarrow q\bar{q}}}{d
x dL}=\frac{2q_{s}^{2}N_{c}\alpha_{em}[x^{2}+(1-x)^{2}]}{M(x)}\mbox{Im}
\int \frac{d\pbt}{(2\pi)^{2}} \,\pbt\cdot \Fb\,,
\label{eq:320}
\eeq
where now $x=E/\omega$, $M(x)=\omega x(1-x)$, and 
$\Fb$ satisfies the integral equation
\beq
\Fb(\pbt)
=
\frac{\pbt}{\pbt^{2}+m_{q}^{2}}
+\frac{2i M(x)}{\pbt^{2}+m_{q}^{2}}
\int \frac{d\qbt}{(2\pi)^{2}}P(\qbt)
[\Fb(\pbt)
-\Fb(\pbt-\qbt)]\,.
\label{eq:330}
\eeq
Substituting (\ref{eq:310}) and (\ref{eq:320}) 
into (\ref{eq:30}) and (\ref{eq:50})
after adding together the bremsstrahlung 
and annihilation contributions one obtains the total photon emission
rate. It can be represented exactly in the form (\ref{eq:10}), (\ref{eq:20})
after the change of variables $\pbt\rightarrow x\pbt$ for the bremsstrahlung,
if one uses the function $\fb$ connected with 
$\Fb$ by the relations: $\fb=-i4E(1-x)\Fb$ for 
bremsstrahlung, and $\fb=-i4\omega x(1-x)\Fb$
for annihilation contributions, respectively.
The difference in the variables for bremsstrahlung 
is connected with the fact that we use the $z$-axis along the initial
quark momentum (which is natural from the point of view of the
light-cone wave function for $q\rightarrow \gamma q$ transition),
and in \cite{AMY1} it is along the photon momentum. In the collinear
approximation both the choices are equivalent.

\vspace{.1cm}
\noindent {\bf 4}.
The proposed derivation of the radiation rate 
is based on the picture of the weakly coupled plasma that is formally
valid at $g\ll 1$. In reality, for the QGP produced in $AA$-collisions,
when $T\lsim (2- 3) T_{c}$, $g\sim 1.5-2$. 
It was proposed that for such temperatures the 
plasma is a strongly coupled QGP (sQGP) \cite{sQGP}.
Nonetheless one can expect that even for the sQGP 
the intuitive physical picture of photon emission will be the same, and
our formulas should be a good approximation (at least for sufficiently
large photon energies) if one somehow 
calculates $P(\r)$ using nonperturbative methods, say,
uses the lattice simulation, or AdS/QCD. 
What is really crucial for our calculations
is the approximation of small parton scattering angles at the scale of the
photon formation length. To understand the situation with this approximation
one can use the results of the analysis of the RHIC
data on jet quenching which give the direct information about $P(\r)$,
and allows one to estimate the $p_{\perp}$ broadening for the thermal
partons.
In terms of $P(\r)$ the $L$-dependence of the $p_{\perp}$ 
distribution can be written as \cite{Z_A2e}
\beq
I(\pbt)=\frac{1}{(2\pi)^{2}}\int d\r \exp[i\pbt\r -P(\r)L]\,.
\label{eq:340}
\eeq
For the purpose of the qualitative estimates one can use the model of
the static Debye screened centers used in \cite{Z_RAA} for 
the analysis of the RHIC data on the nuclear modification factor $R_{AA}$.
Using the oscillator approximation
of the dipole cross section $\sigma(\rho)\approx C\rho^{2}$, one
obtains $\langle p_{\perp}^{2}\rangle=2CnL$. 
Evaluation of $C$ in the double gluon exchange approximation 
with $\alpha_s$ frozen at the value $\alpha_{s}\approx 0.7$
 at low $Q^{2}$ \cite{NZ91,DKT} 
gives 
$C\sim 0.5$ for scattering on a quark. This approximation allows
one to describe the RHIC data on the nuclear modification 
factor \cite{Z_RAA}.
In the oscillator approximation the typical in-medium photon
formation length which accounts for the LPM suppression reads 
$L_{f}^{LPM}\sim [\omega(1-x)/Cnx^{2}]^{1/2}$ \cite{Z_PAN98}. 
Let us consider the final quark which has the smallest energy,
at the scale 
$\sim L_{f}^{LPM}$ one gets
$\langle p_{\perp}^{2}\rangle/p_{z}^{2}\sim
2[{Cn}/{\omega E^{2}(1-x)^{3}}]^{1/2}$. 
For $\omega\sim 6T$ and $E\sim 2\omega$
one obtains $\langle p_{\perp}^{2}\rangle/p_{z}^{2}\sim 0.3$.
With increase of $\omega$ this ratio will be even smaller.
In fact, the angles will be somewhat smaller if one includes
the collisional energy losses which can give the contribution 
about 20-30\% for RHIC conditions. 
Thus, one sees that even though 
the typical angles are not very small, nonetheless,
the collinear approximation seems
to be not far from reality,
at least for sufficiently large photon energies, which, in fact, 
are most interesting from the experimental point of view.
In principle, the fact that the collinearity is not strongly 
violated for RHIC conditions does not contradict to the large value
of $g$. It simply says that the qualitative estimates which just count
the powers of $g$ in the situation when the exact values 
of the coefficients are unknown, are not very reliable. 

\vspace{.1cm}
\noindent{\bf 5}.
In summary, we have proposed a simple alternative derivation of 
the photon emission rate in the pQGP using the LCPI
approach \cite{Z1}. The new derivation is interesting for
several reasons. 
First, the LCPI formulation is convenient for the  
numerical calculations. One can use the methods
previously developed for the case of fast partons \cite{Z_PAN98,Z_RAA}, 
which reduce
the calculation of the radiation rate to solving the two-dimensional
Schr\"odinger equation with
smooth boundary conditions.
Secondly, since the LCPI approach is formulated in the coordinate space,
it can be helpful for clarifying the applicability limits of the
formulas obtained for a uniform plasma when they are applied to the
expanding finite-size plasma produced in $AA$-collisions.
Say, the LCPI formulation can be used to estimate the role
of the boundary effects connected with the initial stage of the QGP evolution
when the radiation rate is the largest. Indeed, in this region, 
say, for the proper
time $\tau\lsim 1$ fm, probability of the photon emission in the partonic
processes may be substantially reduced due to the restriction on 
$|z|\lsim \tau$ in the integral on the right-hand side of (\ref{eq:120}).
This is not possible in the AMY \cite{AMY1} formulation in the momentum
representation.
We leave the  phenomenological applications of our results to 
further publications.

\vspace {.7 cm}
\noindent
{\large\bf Acknowledgements}

\noindent
This research is supported by the "Laboratoire Europ\'een associ\'e
Physique Th\'eorique et Mati\'ere Condens\'ee" (ENS-LANDAU).
BGZ is grateful to the High Energy Group of the ICTP
and the LAPTH for hospitality during his visits there.
The work of BGZ was supported in part by the Grant RFBR
06-02-16078-a.


\newpage

\begin{center}
{\Large \bf Figures}
\end{center}
\begin{figure}[htb]
\begin{center}
\epsfig{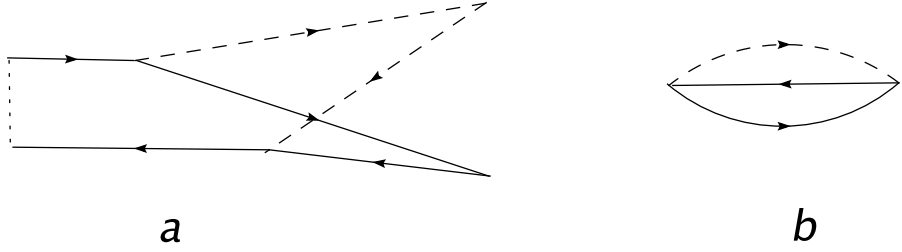}
\end{center}
\caption[.]{
The diagram representation of the $x$-spectrum for 
$q\rightarrow \gamma q$ transition: (a) the starting form in terms 
of the quark and photon propagators,
the left-directed lines correspond to complex conjugate propagators
the vertical dotted line indicates the initial quark density matrix;
(b) the final form corresponding to (\ref{eq:120}) 
in which the only dynamical variable is the transverse
separation between photon and quark.
}
\end{figure}
\begin{figure}[htb] 
\begin{center}
\epsfig{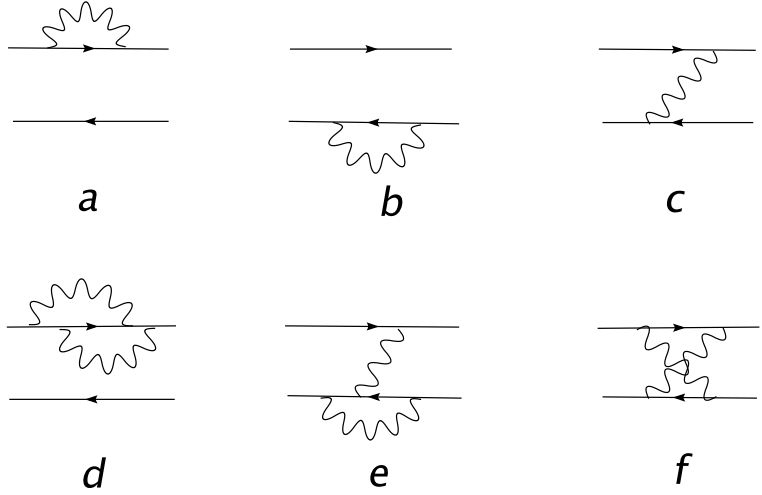}
\end{center}
\caption[.]{
The diagram representation of the function $P(\r)$ in (\ref{eq:150}):
(a),~(b),~(c) the leading order contribution corresponding
 to (\ref{eq:170});
(d),~(e),~(f) some of the diagrams of the order of $g^{4}$.
}
\end{figure}

\end{document}